# DATA-STATIONARY ARCHITECTURE TO EXECUTE QUANTUM ALGORITHMS CLASSICALLY

J. R. Burger
Department of ECE
California State University Northridge
December 9, 2004

*Abstract* – This paper presents a data stationary architecture in which each word has an attached address field. Address fields massively update in parallel to record data interchanges. Words do not move until memory is read for post processing. A sea of such cells can test large-scale quantum algorithms, although other programming is possible.

## 1. INTRODUCTION

Quantum algorithms are better than classical algorithms for certain applications, for example function identification [1, 2]. Quantum algorithms generally have three parts, pre processing, logic implementation, and post processing. Pre processing usually forms a non-sparse state vector whose entries are to be processed in parallel. Logic can be interpreted to mean data interchanges within the state vector as specified by the steps in a 'wiring' diagram [3, 4]. Output processing can use various (real time) digital filtering methods, including Hadamard or Fourier transform.

A quantum algorithm takes a state vector through a sequence of unitary transformations. Circuits are greatly simplified by exploiting the fact that a classical system need not be unitary. That is, a state vector need not be normalized in a classical calculation to achieve a desired result. Another major simplification results by avoiding complex numbers within the state vector. Simply initialize to real integers and perform only real transformations prior to post processing.

To create a system as large as possible in a given technology, it was decided to pre process the state vector prior to loading it into core memory. Core memory represents a state vector. After programmed interchanges within the core, it was decided to perform the post processing outside of the core, where it is easier to do. So core memory does nothing but controlled real integer interchanges as specified by an algorithm, discussed in Section 2 below. Section 3 introduces a concept for practical implementation. Section 4 contains a basic example algorithm. A block diagram is given in Figure 1.

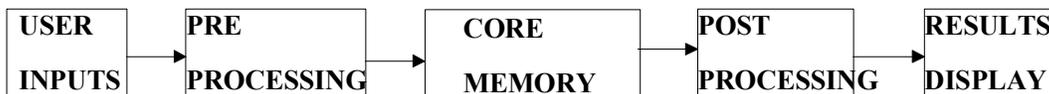

Figure 1. Quantum-inspired Architecture

## 2. CORE MEMORY DESIGN

To show what is needed, an example 'wiring' diagram of a certain quantum algorithm appears in Figure 2.

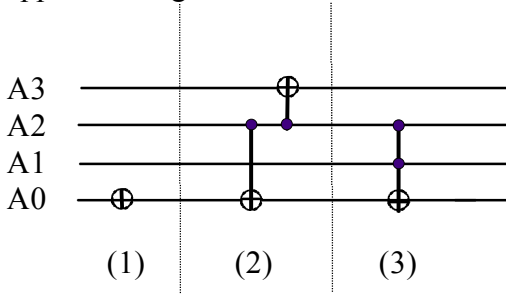

**Figure 2. Wiring Diagram Example**

Unconditional, single controlled NOTs or double controlled NOTs, steps 1, 2 or 3 of Figure 2, can be analyzed as a sequence of interchanges of data within the state vector as in Table 1.

**Table 1 Addresses of Interchanges**: Example: 0011 means M[0011]; UNC = Unconditional; SCN = Single Controlled NOT; DCN = Double Controlled NOT

| Orig. Address | | | SCN Changes | DCN Changes Only |
|---|---|---|---|---|
| | TO=A0 | FM=A2,TO=A0 | FM=A2,TO=A3 | FM=A2,FM=A1,TO=A0 |
| 0000 | 0001 | | | |
| 0001 | 0000 | | | |
| 0010 | 0011 | | | |
| 0011 | 0010 | | | |
| 0100 | 0101 | 0101 | 1100 | |
| 0101 | 0100 | 0100 | 1101 | |
| 0110 | 0111 | 0111 | 1110 | 0111 |
| 0111 | 0110 | 0110 | 1111 | 0110 |
| 1000 | 1001 | | | |
| 1001 | 1000 | | | |
| 1010 | 1011 | | | |
| 1011 | 1010 | | | |
| 1100 | 1101 | 1101 | 0100 | |
| 1101 | 1100 | 1100 | 0101 | |
| 1110 | 1111 | 1111 | 0110 | 1111 |
| 1111 | 1110 | 1110 | 0111 | 1110 |

When $A_k$ is a target, the addresses of the swaps are spaced by $2^k$. This also holds for double controlled NOTs or any transformation 'aimed' at $A_k$

## 3. PRACTICAL CONSIDERATIONS

Data need not physically move. Instead, address fields, attached to each datum in the state vector, can be updated. During output processing it is a simple matter to sort the data into an ordered form, if desired. Each word may be structured as in Figure 3.

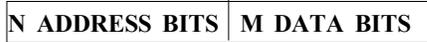

Figure 3. Word Structure

Data needs only M = 2 bits (+1, -1 or 0) after the common normalization factor is taken out. The address field is given only N = 32 bits as an example, although technology supports much larger N. The circuit plan is given in Figure 4.

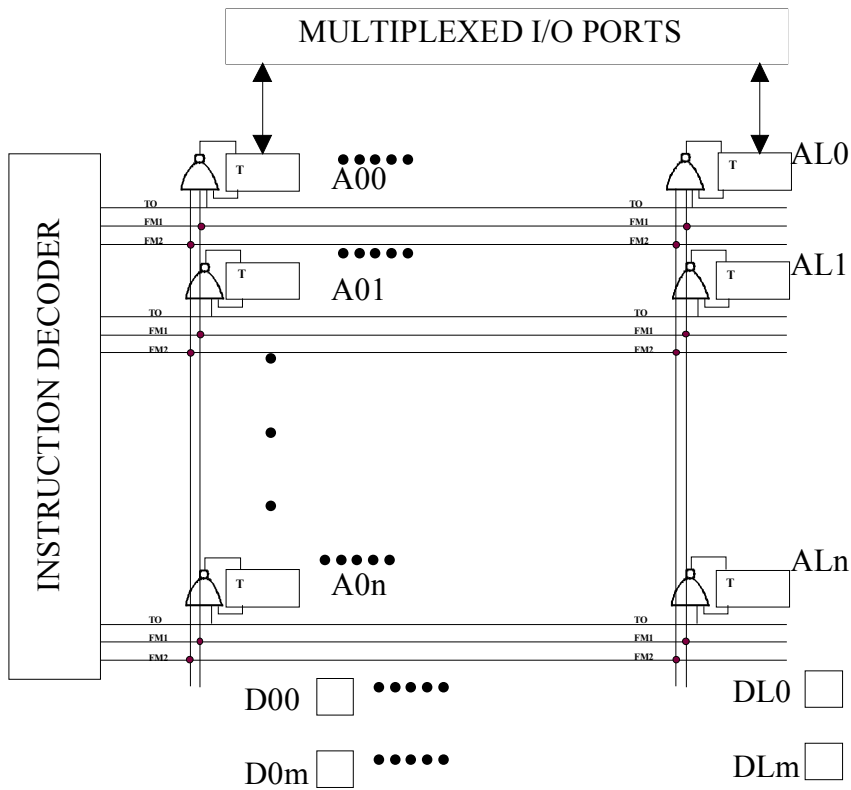

Figure 4. Method of Address Parallel Processing L Words (L = $2^N$)

The vertical buses can be connected to the horizontal buses at the dots using pass transistors. These fire according to the ID (instruction decoder) on the left. For example, consider word 0 address bit 0, that is A00. If TO(A00) = 1 while FM1(A01) = 1 and also FM2(A02) = 1, then the bit A00 will be flipped (implements DCN, Step 3 in Figure 2). This process happens simultaneously for all words, $2^N$ in number, depending on the FM1 and FM2 bits in each individual word. SCN is easily implemented by having FM2 read a hardwired TRUE, leaving FM1 to read the controlling bit of interest.

Size and speed of the processing depend on technology whose performance has been increasing exponentially for many decades now. The following gives a snapshot of today's established technology to see what is possible [5].

Size -- Each word in the above plan takes take 2 transistors for data, and 32 cells for addressing, in which each take 4 transistors; the bus system for each word uses 3 x 32 or 96 transistors; the total is 226 transistors per word.

To estimate number of words, assume a 20 cm radius wafer, and assume that a transistor requires 0.01 $u^2$, where 1 u = 1 x $10^{-6}$ m). Thus 12, 500 x $10^9$ transistors are available in a single wafer. The resulting number of words is over 51 GW (giga words, 1G = $2^{30}$). This corresponds to more than 35 bit of address space, leaving room for decoders at the boundary.

Speed – Distance is assumed 40 cm, so average delay is roughly 320 ns before all buses activate. Any quantum gate for any number of lines in the above architecture can execute in some similar amount of time because of the parallel processing. Post processing time is not included.

### 4. EXAMPLE ALGORITHM

Three lines can be used to identify a binary function f(x) of x = 2 bits. Figure 5 illustrates a certain function. Three lines A2, A1, A0 translate to eight states in a state vector called **y**. Lines A2 and A1 correspond to 4 binary-like codes for the domain of the function. The A0 line 'records' the value of the function. It is possible to answer the following questions about a binary function with only one call to the function.

1) Is the function constant? Constant means it gives either all ones or all zeros as its input counts from 0 to $2^x$-1.
2) Is the function symmetric (or anti-symmetric)? Symmetric means the existence of symmetry in the truth table about the center point. For example, 0110|0110 is symmetric; 0110|1001 is anti-symmetric.
3) Is the function balanced? Balanced means equal numbers of ones and zeros as the input counts through its full range 0 to $2^x$-1.

A classical computer might require up to $2^x$ calls to a function for answering such questions, so the amount of work grows exponentially. In contrast, a quantum computer (or the above architecture) requires only one call to the function. The following example shows how the algorithm works.

Pre Processing -- Assume initialization to (0 1 0 0, 0 0 0 0), that is, a state vector whose short hand notation is $|001\rangle$ . After a transform known as Hadamard transform, the state vector, minus the normalization, reads (1 –1 1 –1, 1 –1 1 –1).

Core Processing -- A certain quantum function involving lines A1 and A2 is implemented.

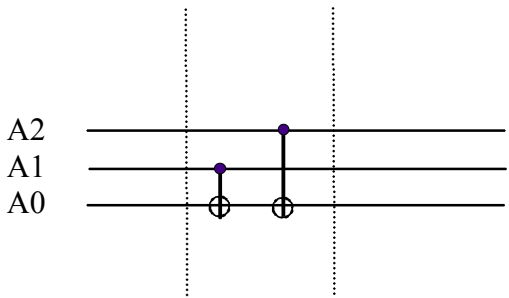

**Figure 5.  Example Quantum Algorithm**

The first SCN transforms the state vector to (1 −1 −1 1, 1 −1 −1 1) as can be seen when addresses re ordered as in Table 2. The second SCN further transforms the state vector to (1 −1 −1 1, -1 1 1 −1) as seen when addresses are placed in order (right side).

**Table 2 XOR Example**

| A(3)a(2)A(1) | y  | Adr / Data | | Re-ordered y | Adr / Data | | Re-ordered y |
|---|---|---|---|---|---|---|---|
| 000 | 1  | 000 | 1  | 1  | 000 | 1  | 1  |
| 001 | -1 | 001 | -1 | -1 | 001 | -1 | -1 |
| 010 | 1  | 011 | 1  | -1 | 011 | 1  | -1 |
| 011 | -1 | 010 | -1 | 1  | 010 | -1 | 1  |
| 100 | 1  | 100 | 1  | 1  | 101 | 1  | -1 |
| 101 | -1 | 101 | -1 | -1 | 100 | -1 | 1  |
| 110 | 1  | 111 | 1  | -1 | 110 | 1  | 1  |
| 111 | -1 | 110 | -1 | 1  | 111 | -1 | -1 |

Post Processing -- Output processing may decode the obvious anti symmetry in the state vector, or one may apply Hadamard transforms to show a basis vector 8 |111⟩, that is, a state vector (0 0 0 0, 0 0 0 8). The 8 implies the state vector is not normalized.

1) It can be reasoned that if the measured output is 8 |001⟩ then the function is indeed constant (zero or one for all input values). Another function, for example the AND function, may have components in |001⟩ but the coefficient will not be 8. A weakness in a physical quantum computer is that in one calculation, it is not certain that the output vector is always |001⟩. That is, it does not calculate the coefficient 8 as a classical calculation does. Consequently, many runs might be required to provide technical confidence in the result.

2) It can be reasoned that if the output is a basis vector other than 8 |001⟩ then the function is either symmetric or anti-symmetric (refer to the above definitions). In this example the result is 8 |111⟩. The function truth table (implied by the wiring diagram) is 01,10; obviously it is symmetric about its center (the comma).

3) Balanced means equal numbers of ones and zeros as the input counts through its full range 0 to $2^x$. Note that the function 01,10 is balanced. Many functions can be balanced, yet neither symmetric nor anti-symmetric. For example, 0111,0001 is balanced, but is NEITHER symmetric nor anti symmetric. A balanced function is distinguished by the fact that it is guaranteed to not have any component in the state |001⟩ (Refer to Deutsch-Josa algorithm [1]). The above system obviously computes in polynomial time what otherwise requires exponential time, as predicted for special cases [6].

**CONCLUSIONS**
The above architecture is mainly a scientific curiosity. Its main purpose is to test larger quantum algorithms using existing technology. Research into quantum algorithms is ongoing [7, 8].

The design presented above is characterized by a very large number of very small registers, each with real signed integers. Each has attached to it a relatively large address field. All addresses can be modified in parallel and conditioned on a logical combination of bits in the address. Once an algorithm finishes, solutions to a problem are established by reading core memory, and by processing to discover symmetries of interest.

This concept was inspired by plans for a quantum computer. Although its applications are currently limited, it could be prove useful someday.